# Ellipsoidal anisotropies in linear elasticity Extension of Saint Venant's work to phenomenological modelling of materials

Ahmad POUYA

LCPC – 58 Bd Lefebvre, 75732, Paris Cedex 15, France

Tel :(+33)1 40 43 52 63, Fax (+33) 1 40 43 65 16

e-mail : pouya@lcpc.fr

**Abstract :** Several families of elastic anisotropies were introduced by Saint Venant (1863) for which the polar diagram of elastic parameters in different directions of the material (indicator surface) are ellipsoidal. These families recover a large variety of models introduced in recent years for damaged materials or as effective modulus of heterogeneous materials. On the other hand, ellipsoidal anisotropy has been used as a guideline in phenomenological modeling of materials. A question that then naturally arises is to know in which conditions the assumption that some indicator surfaces are ellipsoidal allows one to entirely determine the elastic constants. This question has not been rigorously studied in the literature. In this paper, first several basic classes of ellipsoidal anisotropy are presented. Then the problem of determination of the elastic parameters from indicator surfaces is discussed in several basic cases that can occur in phenomenological modelling. Finally the compatibility between the assumption of ellipsoidal form for different indicator surfaces is discussed and in particular, it is shown that if the indicator surfaces of $\sqrt[4]{E(\mathbf{n})}$ and of $\sqrt[-4]{c(\mathbf{n})}$ (where $E(\mathbf{n})$ and $c(\mathbf{n})$ are respectively the Young's modulus and the elastic coefficient in the direction $\mathbf{n}$) are ellipsoidal, then the two ellipsoids have necessarily the same principal axes and the material in this case is orthotropic.

**Keywords** : linear elasticity, anisotropy, indicator surface, ellipsoidal anisotropy, amorphous materials, damaged materials

## 1. Introduction

On the basis of the idea that the material isotropy geometrically corresponds to the image of a sphere, one can naturally seek to extend the isotropic models to anisotropic ones that correspond to an ellipsoidal variation of parameters in different directions. Saint Venant (1863) studied several elasticity models of this type; for instance, a model for which the polar diagram of $\sqrt[4]{E(\mathbf{n})}$, where $E(\mathbf{n})$ is the Young's modulus in direction $\mathbf{n}$, defines an ellipsoidal surface. He regarded these models as being useful for approximation of anisotropic elasticity of amorphous materials.

The models introduced by Saint Venant does not correspond to crystalline types of anisotropy, but recover a large variety of models introduced in recent years for the elasticity tensor of damaged materials (Kachanov 1992, Halm and Dragon 1998, Dragon *et al.* 2000, Chiarelli *et al.* 2003, Alliche 2004) or as effective moduli of heterogeneous media (Milgrom and Shtrikman 1992, Milton 2002). They allows us to represent a three-dimensional anisotropy with reduced number of parameters.

On the other hand the concept of ellipsoidal anisotropy has been naturally used as a guideline for modelling the elasticity of materials, or at least *geomaterials* such as soils, rocks, concrete, etc. The anisotropic behaviour of these materials has been extensively studied in recent years by experimental or numerical methods and is taken into account in geotechnical design (Hefny and LO 1999, Pouya and Reiffsteck 2003) and also in the study of seismic wave propagation, genesis of geological structures (Pan and Amadei 1996), micro-cracking of rocks (Takemura *et al.* 2003), etc. Most of times, a rough representation of the anisotropy with a minimum number of parameters is sufficient for the purposes of these studies. Peres Rodrigues (1970) tried to fit the measured values of the Young's modulus in different direction for several varieties of rocks by ellipsoids. For the study of seismic wave propagation in geological layers, Daley and Hron (1979) defined the "elliptically anisotropic" medium as being characterized by elliptical P wave fronts emanating from a point source. This concept was widely used in geophysical studies and examined by Thomsen (1986) in the context of "weak anisotropy" and transversal isotropy for a large variety of sedimentary rocks. Louis *et al.* (2004) proposed a simplified method to analyse the P-way velocity data in anisotropic rocks which supposes implicitly an ellipsoidal approximation of some elastic parameters. Pouya and Reiffsetck (2003) remarked that some Bohler's (1975) data on the Young's Modulus of different soils presents an ellipsoidal property, and showed that this assumption allows to simplify the modelling of foundations. As a mater of fact, it was shown by Pouya(2000) and Pouya and Zaoui (2005) that many closed form solutions for basic problems in linear isotropic materials can be extended by a *linear transformation* to a variety "ellipsoidal" materials.

The concept of ellipsoidal anisotropy in elasticity thus seems an attractive guideline for phenomenological modelling of amorphous, micro-cracked or damaged materials since it simplifies data analysis and defines models with reduced number of parameters and interesting theoretical properties. Nevertheless, some theoretical questions concerning the existence and uniqueness of an elasticity tensor solution for one or more ellipsoidal indicator surfaces have not been rigorously examined in the literature. For instance, contrarily to what was supposed by Peres Rodrigues (1970), the indicator surface of $E(\mathbf{n})$ can never be an ellipsoid (different from a sphere), and so the parameters fitted by this author do not define an elasticity tensor. In this paper we first present some elastic parameters which can have ellipsoidal variation. Then the problem of determination of the elasticity tensor from indicator surfaces, which was studied by Hé and Curnier (1995) in restricted cases of damaged materials, will be examined for basic cases of ellipsoidal surfaces.

## 2. Notation

In what follows, light-face (Greek or Latin) letters denote scalars; bold-face minuscules and majuscules designate respectively vectors and second rank tensors or double-index matrices; outline letters are reserved for fourth rank tensors. The convention of summation on repeated indices is used implicitly. The scalar product of two vectors is labelled as $\mathbf{a.b} = a_i b_i$, and their *symmetric* tensor product as $\mathbf{a}\otimes\mathbf{b}$ with $(\mathbf{a}\otimes\mathbf{b})_{ij} = (a_i b_j + a_j b_i)/2$. The matrix product is labelled as **AB**, and the inner product as $\mathbf{A:B} = A_{ij}B_{ij}$. The operation of the fourth rank tensor $\mathbb{C}$ on **A** will be labelled by $\mathbb{C}\mathbf{:A}$ with $(\mathbb{C}\mathbf{:A})_{ij} = C_{ijkl}A_{kl}$ and the operation of **A** on **a,** by **A.a**.
For an elasticity tensor $\mathbb{C}$ verifying the symmetries $C_{ijkl} = C_{ijlk} = C_{klij}$, two distinct double index notations are introduced : the double sub-stript (ij) is first abbreviated to a single sub-script ($\alpha$) running from 1 to 6 by the following rule :

$$11 \to 1,\quad 22 \to 2,\quad 33 \to 3,\quad 23 \to 4,\quad 13 \to 5,\quad 12 \to 6$$

The *matrix notation* ***C*** is defined by its components $c_{\alpha\beta} = C_{(ij)(kl)}$, and the *dual matrix notation* $\underline{\boldsymbol{C}}$ for the same tensor $\mathbb{C}$, by its components $\underline{c}_{\alpha\beta} = \kappa\, c_{\alpha\beta}$ with :

$$\kappa = 1 \text{ if } \alpha \leq 3 \text{ and } \beta \leq 3$$
$$\kappa = 4 \text{ if } \alpha > 3 \text{ and } \beta > 3$$
$$\kappa = 2 \textit{ elsewhere}$$

Let us notice that the elastic compliances that are commonly designated in the literature by $s_{\alpha\beta}$ (Lekhnitskii 1963, Sirotine and Chaskolskaia 1984, Ting 1996), are here designated by $\underline{s}_{\alpha\beta}$.

## 3. Fourth order indicator surfaces

The *indicator surface* for a "mono-directional" elastic parameter, *i.e.*, a parameter depending upon the elasticity tensor $\mathbb{C}$ and only one direction **n**, is the polar diagram $\mathbf{x} = r(\mathbf{n})\,\mathbf{n}$ where **n** is a unit vector and r(**n**) is the value of the elastic parameter in the direction **n**. Some examples of "mono-directional" parameters are the *Young's modulus*, *bulk modulus*, *elastic coefficient* and *hydrostatic coefficient in direction* **n** defined respectively by :

$$E(\mathbf{n}) = [(\mathbf{n}\otimes\mathbf{n})\mathbf{:}\mathbb{S}\mathbf{:}(\mathbf{n}\otimes\mathbf{n})]^{-1} \qquad (1)$$
$$b(\mathbf{n}) = [\delta\mathbf{:}\mathbb{S}\mathbf{:}(\mathbf{n}\otimes\mathbf{n})]^{-1} \qquad (2)$$
$$c(\mathbf{n}) = (\mathbf{n}\otimes\mathbf{n})\mathbf{:}\mathbb{C}\mathbf{:}(\mathbf{n}\otimes\mathbf{n}) \qquad (3)$$
$$h(\mathbf{n}) = \delta\mathbf{:}\mathbb{C}\mathbf{:}(\mathbf{n}\otimes\mathbf{n}) \qquad (4)$$

where $\delta$ is the second-order unit tensor and $\mathbb{S} = \mathbb{C}^{-1}$. Other mono-directional parameters, such as the *Poisson's ratio in direction* **n**, $\nu(\mathbf{n}) = [1 - E(\mathbf{n})/b(\mathbf{n})]/2$, or the *torsion modulus in direction* **n**, $\tau(\mathbf{n}) = [2(\delta_{ik} - n_i n_k)S_{ijkl}n_j n_l]^{-1}$, have also been introduced by some authors and their indicator surfaces have been studied (Sirotine and Chaskolskaia 1984). We limit our investigation to the study of the parameters defined by the relations (1) to (4), which are the basic parameters that can be experimentally determined from simple traction or extension tests.

The indicator surface of the Young's modulus has been widely studied for all types of materials. The polar equation of this surface :

$$r(\mathbf{n}) = E(\mathbf{n}) = [(\mathbf{n}\otimes\mathbf{n}):\mathbb{S}:(\mathbf{n}\otimes\mathbf{n})]^{-1} \quad (5)$$

can be transformed, by using $r^2 = \mathbf{x}.\mathbf{x}$ and $\mathbf{n} = \mathbf{x}/r$, into the following polynomial equation:

$$[(\mathbf{x}\otimes\mathbf{x}):\mathbb{S}:(\mathbf{x}\otimes\mathbf{x})]^2 = (\mathbf{x}.\mathbf{x})^3 \quad (6)$$

This $8^{th}$ order surface has been in particular deeply investigated by Cazzani and Marco (2003, 2005) for cubic, hexagonal and tetragonal symmetries. It takes in general very complex forms (see Figure 1) and it can be shown that it can not be an ellipsoid different from a sphere (Appendix 1). So this surface is not suitable for an ellipsoidal approximation of E(**n**) values contrarily to that what supposed by Peres Rodrigues (1970). On the other hand, if the indicator surface of $\sqrt[4]{E(\mathbf{n})}$ is considered then a fourth-order surface is found which is described by the following equation :

$$(\mathbf{x}\otimes\mathbf{x}):\mathbb{S}:(\mathbf{x}\otimes\mathbf{x}) = 1 \quad (7)$$

This surface, which contains exactly the same information that the previous one, *degenerates* for some cases of materials to second-order surfaces, and more precisely, to ellipsoids. This will define a class of materials that will be said to have a variety of *ellipsoidal anisotropy*.

Other fourth order indicator surfaces which can degenerate to ellipsoids are those of $\sqrt{E(\mathbf{n})}$, $[c(\mathbf{n})]^{-1/4}$ and $[c(\mathbf{n})]^{-1/2}$. The indicator surfaces of $[b(\mathbf{n})]^{1/2}$ and $[h(\mathbf{n})]^{-1/2}$ are always ellipsoidal. The denomination and equations of these surfaces are given in the Table (1).

| Indicator surface | Elastic parameter | Equation |
|---|---|---|
| $F_4(\mathbb{C})$ | $[(\mathbf{n}\otimes\mathbf{n}):\mathbb{C}:(\mathbf{n}\otimes\mathbf{n})]^{-1/4}$ | $(\mathbf{x}\otimes\mathbf{x}):\mathbb{C}:(\mathbf{x}\otimes\mathbf{x}) = 1$ |
| $F_2(\mathbb{C})$ | $[(\mathbf{n}\otimes\mathbf{n}):\mathbb{C}:(\mathbf{n}\otimes\mathbf{n})]^{-1/2}$ | $(\mathbf{x}\otimes\mathbf{x}):\mathbb{C}:(\mathbf{x}\otimes\mathbf{x}) = \mathbf{x}.\mathbf{x}$ |
| $G_4(\mathbb{C})$ | $[(\mathbf{n}\otimes\mathbf{n}):\mathbb{S}:(\mathbf{n}\otimes\mathbf{n})]^{-1/4}$ | $(\mathbf{x}\otimes\mathbf{x}):\mathbb{S}:(\mathbf{x}\otimes\mathbf{x}) = 1$ |
| $G_2(\mathbb{C})$ | $[(\mathbf{n}\otimes\mathbf{n}):\mathbb{S}:(\mathbf{n}\otimes\mathbf{n})]^{-1/2}$ | $(\mathbf{x}\otimes\mathbf{x}):\mathbb{S}:(\mathbf{x}\otimes\mathbf{x}) = \mathbf{x}.\mathbf{x}$ |
| $f(\mathbb{C})$ | $[\boldsymbol{\delta}:\mathbb{C}:(\mathbf{n}\otimes\mathbf{n})]^{-1/2}$ | $\boldsymbol{\delta}:\mathbb{C}:(\mathbf{x}\otimes\mathbf{x}) = 1$ |
| $g(\mathbb{C})$ | $[\boldsymbol{\delta}:\mathbb{S}:(\mathbf{n}\otimes\mathbf{n})]^{-1/2}$ | $\boldsymbol{\delta}:\mathbb{S}:(\mathbf{x}\otimes\mathbf{x}) = 1$ |

Table 1 : Indicator surfaces of different elastic parameters and their polynomial equation

## 4. Saint Venant's anisotropies

Saint Venant (1863) introduced several families of orthotropic materials for which one or more of the surfaces $F_4(\mathbb{C})$, $G_4(\mathbb{C})$, $F_2(\mathbb{C})$ or $G_2(\mathbb{C})$ is ellipsoidal. Let us note by $\mathcal{B} = \{\mathbf{e}_1, \mathbf{e}_2, \mathbf{e}_2\}$ a set of unit vectors defining a Cartesian coordinate system, and by $\Omega(\mathcal{B})$, the family of orthotropic materials with planes of symmetry given by $\mathcal{B}$.

The first family defined by Saint Venant is the sub-set of $\Omega(\mathcal{B})$, denoted here by $\Phi_4(\mathcal{B})$, for which the following relations between the parameters are satisfied :

$$c_{44} = \frac{\sqrt{c_{22}\ c_{33}} - c_{23}}{2}, \qquad c_{55} = \frac{\sqrt{c_{11}\ c_{33}} - c_{13}}{2}, \qquad c_{66} = \frac{\sqrt{c_{11}\ c_{22}} - c_{12}}{2} \quad (8)$$

For $\mathbb{C}\in\Phi_4(\mathcal{B})$, the equation of $F_4(\mathbb{C})$, given in the Table (1), reduces to the equation of an ellipsoid with principal axes $\mathcal{B}$ and semi-diameters $\{(c_{11})^{-1/4}, (c_{22})^{-1/4}, (c_{33})^{-1/4}\}$:

$$\sqrt{c_{11}}\ x_1^2 + \sqrt{c_{22}}\ x_2^2 + \sqrt{c_{33}}\ x_3^2 = 1 \tag{9}$$

The relations (8) were considered by many authors, in the context of transversal isotropy, as defining a *degenerate* case (Pan and Chou 1976).

The second family is that of orthotropic tensors for which relations analogous to (8) hold between the parameters $s_{\alpha\beta}$. In terms of $\underline{s}_{\alpha\beta}$, these relations become :

$$\underline{s}_{44} = 2\left(\sqrt{\underline{s}_{22}\,\underline{s}_{33}} - \underline{s}_{23}\right), \quad \underline{s}_{55} = 2\left(\sqrt{\underline{s}_{11}\,\underline{s}_{33}} - \underline{s}_{13}\right), \quad \underline{s}_{66} = 2\left(\sqrt{\underline{s}_{11}\,\underline{s}_{22}} - \underline{s}_{12}\right) \tag{10}$$

and in terms of elastic modulus and Poissons's ratio (see the expression 35), they become :

$$G_{23} = \frac{\sqrt{E_2 E_3}}{2\left(1+\sqrt{\nu_{23}\nu_{32}}\right)}, \quad G_{31} = \frac{\sqrt{E_3 E_1}}{2\left(1+\sqrt{\nu_{31}\nu_{13}}\right)}, \quad G_{12} = \frac{\sqrt{E_1 E_2}}{2\left(1+\sqrt{\nu_{12}\nu_{21}}\right)} \tag{11}$$

This family, which was quoted also by Lekhnitskii (1963), will be labelled here $\Gamma_4(\mathcal{B})$. For $\mathbb{C}\in\Gamma_4(\mathcal{B})$ the surface $G_4(\mathbb{C})$ is an ellipsoid with principal axes $\mathcal{B}$ and semi-diameters $\{(s_{11})^{-1/4}, (s_{22})^{-1/4}, (s_{33})^{-1/4}\}$, or $\{\sqrt[4]{E_1}, \sqrt[4]{E_2}, \sqrt[4]{E_3}\}$. In the context of transversal isotropy (with axis $x_3$), the third equality of (10) is always satisfied and the two first one are equivalent. In this context, these relations were considered by Boehler (1982) as defining a case of *limited anisotropy* with elliptical properties.

The elements of $\Phi_4(\mathcal{B})$ and $\Gamma_4(\mathcal{B})$ depend on 6 intrinsic parameters. The intersection between $\Phi_4(\mathcal{B})$ and $\Gamma_4(\mathcal{B})$ is denoted by $\Psi(\mathcal{B})$ :

$$\Psi(\mathcal{B}) = \Phi_4(\mathcal{B}) \cap \Gamma_4(\mathcal{B}) \tag{12}$$

This family of materials, as showed by Saint Venant, depends on 4 intrinsic parameters in the coordinates system $\mathcal{B}$, and can be represented equivalently by one of the following expressions of $\boldsymbol{C}$ or of $\underline{\boldsymbol{S}}$ in this coordinates system :

$$\Psi(\mathcal{B}) : \qquad \boldsymbol{C} = \begin{bmatrix} c_{11} & \eta\sqrt{c_{11}\,c_{22}} & \eta\sqrt{c_{11}\,c_{33}} & & & \\ & c_{22} & \eta\sqrt{c_{22}\,c_{33}} & & & \\ & & c_{33} & & & \\ & & & \frac{1-\eta}{2}\sqrt{c_{22}\,c_{33}} & & \\ & & & & \frac{1-\eta}{2}\sqrt{c_{11}\,c_{33}} & \\ & & & & & \frac{1-\eta}{2}\sqrt{c_{11}\,c_{22}} \end{bmatrix} \tag{13}$$

$$\Psi\ (\mathcal{B}): \qquad \underline{\mathbf{S}} = \begin{bmatrix} \frac{1}{E_1} & \frac{-\nu}{\sqrt{E_1E_2}} & \frac{-\nu}{\sqrt{E_1E_3}} & & & \\ \frac{-\nu}{\sqrt{E_1E_2}} & \frac{1}{E_2} & \frac{-\nu}{\sqrt{E_2E_3}} & & & \\ \frac{-\nu}{\sqrt{E_1E_3}} & \frac{-\nu}{\sqrt{E_2E_3}} & \frac{1}{E_3} & & & \\ & & & \frac{2(1+\nu)}{\sqrt{E_2E_3}} & & \\ & & & & \frac{2(1+\nu)}{\sqrt{E_3E_1}} & \\ & & & & & \frac{2(1+\nu)}{\sqrt{E_1E_2}} \end{bmatrix} \qquad (14)$$

The two sets of parameters ($c_{11}$, $c_{22}$, $c_{33}$, $\eta$) and ($E_1$, $E_2$, $E_3$, $\nu$) are related by :

$$\nu = \frac{\eta}{1+\eta}, \quad E_\alpha = \frac{(1-\eta)(1+2\eta)}{1+\eta} c_{\alpha\alpha} \quad (\alpha=1,2,3;\ \text{non summation upon } \alpha) \qquad (15)$$

Pouya and Zaoui (2005) showed that for this family of materials, $\mathbb{C}$ can be written in the following forms :

$$C_{ijkl} = P_{im} P_{jn} P_{kp} P_{lq} \tilde{C}_{mnpq} \qquad (16)$$

$$C_{ijkl} = \lambda D_{ij} D_{kl} + \mu\, (D_{ik} D_{jl} + D_{il} D_{jk}) \qquad (17)$$

In these relations $\tilde{\mathbb{C}}$ represents an isotropic elasticity tensor with Lamé constants $\lambda=\eta$ and $\mu=(1-\eta)/2$, **P** is given in the coordinates system $\mathcal{B}$ by :

$$\mathbf{P} = \text{diag}\left(\sqrt[4]{c_{11}},\ \sqrt[4]{c_{22}},\ \sqrt[4]{c_{33}}\right) \qquad (18)$$

and **D** = **PP**.

Conversely, it can be checked that for the materials defined by (17), $F_4(\mathbb{C})$ and $G_4(\mathbb{C})$ both are ellipsoidal. Therefore (17) defines $\Psi$ type materials where :

$$\Psi = \bigcup_{\mathcal{B}} \Psi(\mathcal{B}) \qquad (19)$$

The expression (17) has been used in some micromechanical studies as representing the effective modulus of heterogeneous media (Milgrom & Shtrikman 1992, Milton 2002). $\Psi$ type materials present interesting theoretical properties. Saint Venant showed that D'Alembert's displacement potentials as well as the solution for plane waves propagation in isotropic elasticity can be extended to this family. Pouya (2000) and Pouya and Zaoui (2005) showed that the relation (16) between the two elasticity tensors allows one to extend many closed form solutions for basic problems of elasticity to $\Psi$ type materials. Some examples of results extended in this way to $\Psi$ type materials are : the Eshelby tensor for inclusion-matrix problem (Milgrom & Shtrikman 1992, Pouya 2000), Green function for infinite space (Pouya 2000) and for half-space (Pouya and Zaoui 2005). Extension of the Green function solutions for two joined semi-infinite isotropic solids (Rongved 1955) or for layerd medium constituted of isotropic materials (Benitez & Rosakis 1987) to solids constituted of $\Psi$ type materials would also be possible (Pouya and Zaoui 2005).

Thus theoretical simplifications can result from choosing $\Psi$ type models for representing (as an approximation) the elastic anisotropy of materials.

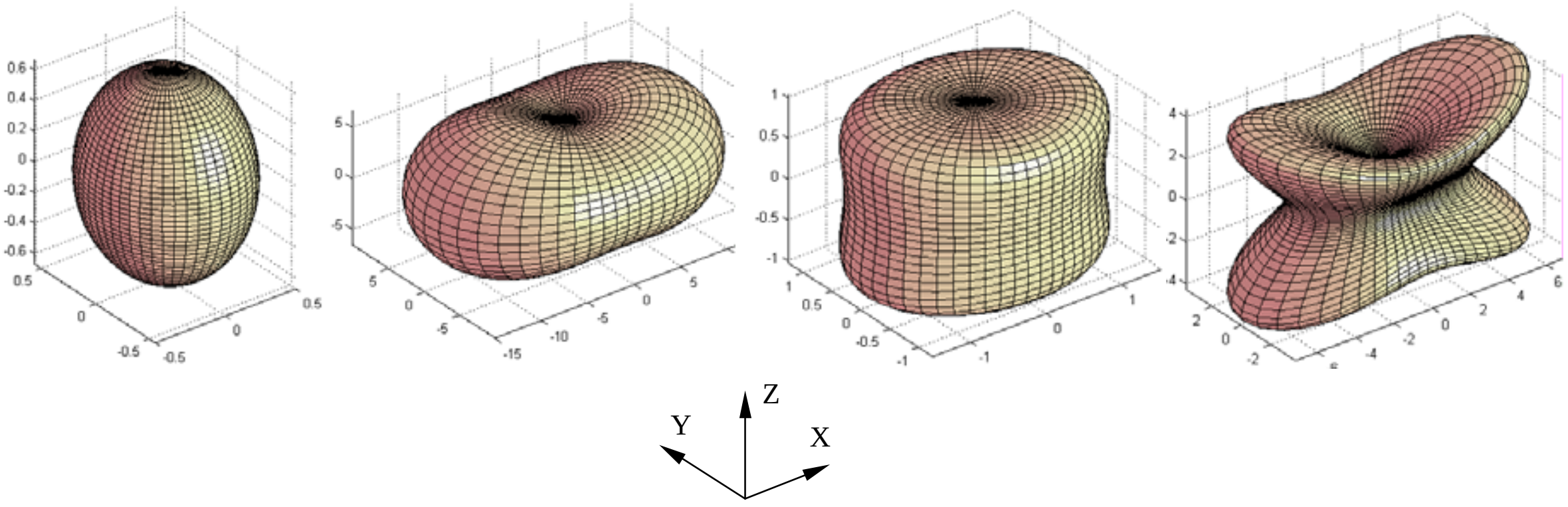

(a) (b) (c) (d)

Figure 1 : Indicator surfaces for an element of $\Phi_4$

For $\mathbb{C}\in\Phi_4(\mathcal{B})$ with $c_{11}$ = 15, $c_{22}$ = 10, $c_{33}$ = 5, $c_{12}$ = -3, $c_{13}$ = 4, $c_{23}$ = 4.5, and $c_{44}$, $c_{55}$, $c_{66}$ given by the relation (8), the indicator surfaces of $\sqrt[-4]{c(\mathbf{n})}$ , $c(\mathbf{n})$ , $\sqrt[4]{E(\mathbf{n})}$ and $E(\mathbf{n})$ are respectively an ellipsoid (a), a 10$^{th}$ order (b), 4$^{th}$ order (c) and 8$^{th}$ order surface (d).

Let us now consider the model (17). The tensor **D** can be decomposed as $\mathbf{D} = p\,\delta + \alpha\mathbf{G}$ where $p = (\mathbf{D}:\delta)/3$ and **G** is traceless and normalized to unity: $\mathbf{G}:\delta = 0$, $\mathbf{G}:\mathbf{G}=1$. The parameter p can be chosen with $\lambda$ and $\mu$: we can take p =1 without loss of generality and write $\mathbf{D} = \delta + \alpha\mathbf{G}$. If $\alpha = 0$, then (17) gives the isotropic elasticity tensor. In the context of "weak anisotropy", *i.e.* when $|\alpha|<<1$, the first order expansion of (17) with respect to $\alpha$ leads to:

$$C_{ijkl} = \lambda\delta_{ij}\delta_{kl} + \mu\,(\delta_{ik}\delta_{jl} + \delta_{il}\delta_{jk}) + a_1(\delta_{ij}G_{kl} + \delta_{kl}G_{ij}) + a_2(\delta_{ik}G_{jl} + \delta_{il}G_{jk} + \delta_{jl}G_{ik} + \delta_{jk}G_{il}) \quad (20)$$

with $a_1 = \lambda\alpha$ and $a_2 = \mu\alpha$. The expression (20) with independent values for $a_1$ and $a_2$ (and not necessarily infinitesimal) has been widely used in the literature for representing the elasticity tensor of damaged materials. It has been obtained by Kachanov (1992) as the effective moduli of micro-cracked media and then widely used as a phenomenological model for damaged geomaterials (Chiarelli *et al.* 2000, Alliche 2004) or as an intermediary between micromechanical and phenomenological models for further theoretical investigations (Halm and Dragon 1998, Dragon *et al.* 2000).

It is interesting to notice that the model (20) can be defined directly by an ellipsoidal property: for this model the surface $F_2(\mathbb{C})$ is ellipsoidal. Saint Venant defined a sub set of $\Omega(\mathcal{B})$, denoted here by $\Phi_2(\mathcal{B})$, for which the following relations are satisfied between the parameters:

$$c_{44} = \frac{c_{22} + c_{33} - 2c_{23}}{4}, \quad c_{55} = \frac{c_{11} + c_{33} - 2c_{13}}{4}, \quad c_{66} = \frac{c_{11} + c_{22} - 2c_{12}}{4} \tag{21}$$

For $\mathbb{C} \in \Phi_2(\mathcal{B})$, the surface $F_2(\mathbb{C})$ is an ellipsoid with principal axes $\mathcal{B}$ and semi-diameters $(1/\sqrt{c_{11}}, 1/\sqrt{c_{22}}, 1/\sqrt{c_{33}})$. The elements of this family have the expression (20) where **G** is diagonal in $\mathcal{B}$. It can be checked that the expression (20) defines exactly the family $\Phi_2$ of materials given by:

$$\Phi_2 = \bigcup_{\mathcal{B}} \Phi_2(\mathcal{B}) \tag{22}$$

Finally, a sub family of $\Omega(\mathcal{B})$, denoted here by $\Gamma_2(\mathcal{B})$, was defined by Saint Venant for which a (21) type relation is verified between $s_{\alpha\beta}$ parameters, or equivalently (see also Lekhnitskii 1963):

$$\frac{1}{G_{23}} - \frac{2\sqrt{\nu_{23}\nu_{32}}}{\sqrt{E_2 E_3}} = \frac{1}{E_2} + \frac{1}{E_3}, \quad \frac{1}{G_{31}} - \frac{2\sqrt{\nu_{31}\nu_{13}}}{\sqrt{E_3 E_1}} = \frac{1}{E_3} + \frac{1}{E_1}, \quad \frac{1}{G_{12}} - \frac{2\sqrt{\nu_{12}\nu_{21}}}{\sqrt{E_1 E_2}} = \frac{1}{E_1} + \frac{1}{E_2} \tag{23}$$

For this family, $G_2(\mathbb{C})$ is ellipsoidal with semi-diameters $\{\sqrt{E_1}, \sqrt{E_2}, \sqrt{E_3}\}$.

## 5. Characterization of the material by indicator surfaces

For the families studied here above, the expression of the tensors $\mathbb{C}$ or $\mathbb{S}$ is given *a priori*. Let us suppose now that the values of some mono-directional elastic parameters have been determined in different directions by experimental or numerical methods. E(**n**) values have been frequently determined experimentally by coring samples in different directions in rocks and soils (Boehler 1975). Numerical homogenisation methods allows one to easily determine c(**n**) or E(**n**) in different directions. For instance, Min and Jing (2003) determined the E(**n**) values for a fractured rock mass by applying a compression parallel to the sides of a square REV and by rotating the REV with respect to the fractures (Figure 2). Acoustic measurements are widely used for determining the parameters of anisotropic elasticity in different directions. Sometimes also a complete set of numerical values is determined by this method for the 21 parameters of general anisotropy (Homand *et al.* 1993). In this case also it can be interesting to seek for an approximate model having a reduced number of parameters and the study of the indicator surfaces can constitute a good guideline for this purpose.

Consider that the polar diagrams of $[c(\mathbf{n})]^{-1/4}$, $\sqrt[4]{E(\mathbf{n})}$, $[c(\mathbf{n})]^{-1/2}$ or $\sqrt{E(\mathbf{n})}$ have been constructed for one on several samples (or REVs) by experimental (or numerical) methods. Let us suppose that they can be sufficiently well fitted by ellipsoidal surfaces (Figure 3). The problem then is to know how to deduce $\mathbb{C}$ from the fitting parameters. A preliminary question would be to know if any pair of ellipsoids fitted for two different parameters is compatible with the existence of an elasticity tensor. To study these problems we first characterize the classes of materials corresponding to one condition of ellipsoidal indicator surface, and then we will study the intersection between two different classes.

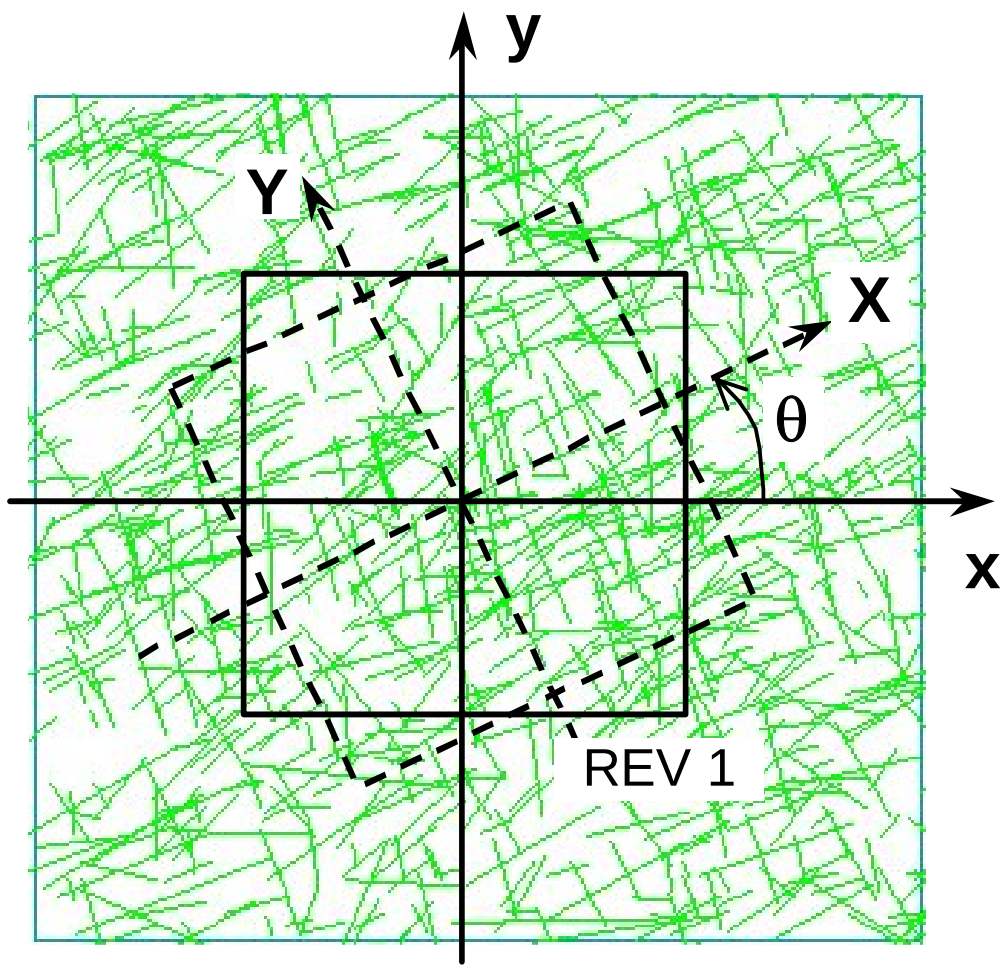

Figure 2 : Numerical determination of the elastic parameters in different directions of a heterogeneous medium by rotation of the REV with respect to the medium

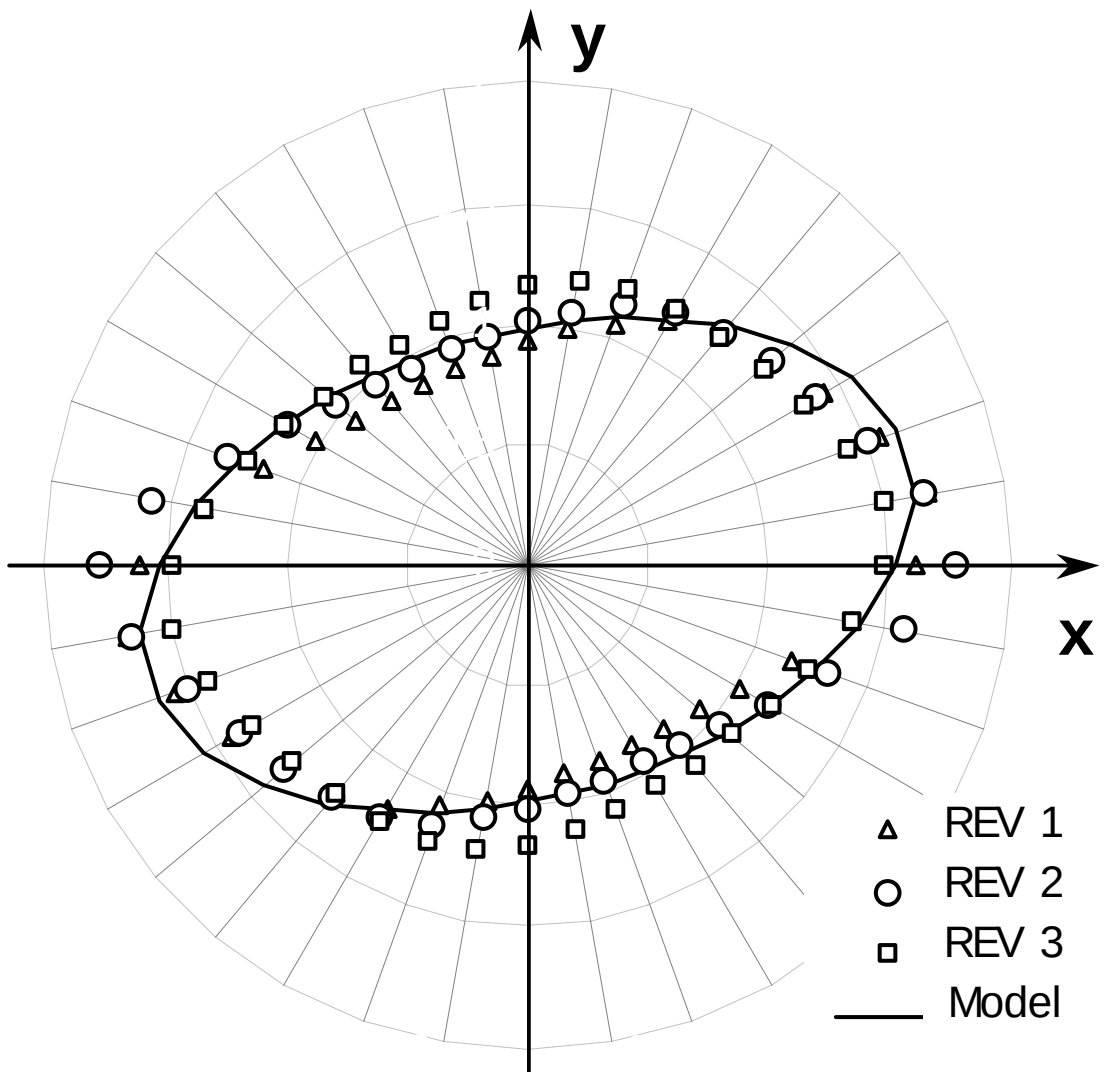

Figure 3 : Fitting of the experimental or numerical indicator surfaces obtained for one or several samples by ellipsoids

### 5.1. Material classes

We define $\hat{\Phi}_4(\mathcal{B})$ as the class of tensors $\mathbb{C}$ for which $F_4(\mathbb{C})$ is an ellipsoid admitting the system $\mathcal{B}$ as principal axes. This class obviously includes $\Phi_4(\mathcal{B})$, but, as it will be seen further, is larger than $\Phi_4(\mathcal{B})$. In the same way, the classes $\hat{\Phi}_2(\mathcal{B})$, $\hat{\Gamma}_4(\mathcal{B})$, $\hat{\Gamma}_2(\mathcal{B})$, $\hat{\varphi}(\mathcal{B})$ and $\hat{\gamma}(\mathcal{B})$ are defined as the classes of tensors $\mathbb{C}$ for which respectively $F_2(\mathbb{C})$, $G_4(\mathbb{C})$, $G_2(\mathbb{C})$, $f(\mathbb{C})$ and $g(\mathbb{C})$ is ellipsoidal with principal axes $\mathcal{B}$. The classes $\hat{\Phi}_4$, $\hat{\Phi}_2$, $\hat{\Gamma}_4$ and $\hat{\Gamma}_2$ are defined as the sets of tensors $\mathbb{C}$ for which the surfaces $F_4(\mathbb{C})$, $F_2(\mathbb{C})$, $G_4(\mathbb{C})$ and $G_2(\mathbb{C})$ respectively are ellipsoidal :

$$\hat{\Phi}_4 = \bigcup_{\mathcal{B}} \hat{\Phi}_4(\mathcal{B}) , \qquad \hat{\Phi}_2 = \bigcup_{\mathcal{B}} \hat{\Phi}_2(\mathcal{B}) , \qquad \hat{\Gamma}_4 = \bigcup_{\mathcal{B}} \hat{\Gamma}_4(\mathcal{B}) , \qquad \hat{\Gamma}_2 = \bigcup_{\mathcal{B}} \hat{\Gamma}_2(\mathcal{B}) \qquad (24)$$

The spherical classes $\hat{\Phi}_s$, $\hat{\Gamma}_s$ , $\hat{\varphi}_s$ and $\hat{\gamma}_s$ are defined as the sets of tensors $\mathbb{C}$ for which respectively the surfaces $F_4(\mathbb{C})$ (or $F_2(\mathbb{C})$), $G_4(\mathbb{C})$ (or $G_2(\mathbb{C})$), $f(\mathbb{C})$ and $g(\mathbb{C})$ are spherical. Finally $I_s$ represents as the class of isotropic tensors.

Two different indicator surfaces at least are required to fully determine $\mathbb{C}$. Hé and Curnier (1995) showed that if the indicator surfaces of the parameters c(**n**) and h(**n**) defined by (3) and (4) are spherical, then $\mathbb{C}$ is isotropic. This is equivalent to say that if the surfaces $F_4(\mathbb{C})$ and $f(\mathbb{C})$ are spherical, then $\mathbb{C}$ is isotropic :

$$\hat{\Phi}_s \cap \hat{\varphi}_s = I_s \qquad (25)$$

Equivalently, if $G_4(\mathbb{C})$ and $g(\mathbb{C})$ are spherical, then $\mathbb{C}$ is isotropic : $\hat{\Gamma}_s \cap \hat{\gamma}_s = I_s$.

These results will be extended in the following to ellipsoidal surfaces. Two cases will be studied : when $F_4(\mathbb{C})$ or $F_2(\mathbb{C})$ is ellipsoidal with the same planes of symmetry than $f(\mathbb{C})$, and when $F_4(\mathbb{C})$ and $G_4(\mathbb{C})$ are both ellipsoidal.

### 5.2. Expression of the elasticity tensor for different classes

Let us first determine the explicit expressions of tensors $\mathbb{C}$ or $\mathbb{S}$ for the classes introduced above. For $\mathbb{C} \in \hat{\Phi}_4(\mathcal{B})$, the surface $F_4(\mathbb{C})$ is an ellipsoid with principal axes $\mathcal{B}$. If ($D_1$, $D_2$, $D_3$) denote the semi-diameters of this ellipsoid then its equation in the coordinate system $\mathcal{B}$ is :

$$\frac{x_1^2}{D_1^2} + \frac{x_2^2}{D_2^2} + \frac{x_3^2}{D_3^2} = 1 \tag{26}$$

This equation must be equivalent to that given in the Table (1) for $F_4(\mathbb{C})$:

$$\forall\, \mathbf{x} \;; \qquad \frac{x_1^2}{D_1^2} + \frac{x_2^2}{D_2^2} + \frac{x_3^2}{D_3^2} = 1 \;\Leftrightarrow\; (\mathbf{x}\otimes\mathbf{x}):\mathbb{C}:(\mathbf{x}\otimes\mathbf{x}) = 1 \tag{27}$$

By using :

$(\mathbf{x}\otimes\mathbf{x}):\mathbb{C}:(\mathbf{x}\otimes\mathbf{x}) =$

$$\begin{aligned} & c_{11}\, x_1^4 + c_{22}\, x_2^4 + c_{33}\, x_3^4 + (2c_{23} + 4c_{44})\, x_2^2 x_3^2 + (2c_{31} + 4c_{55})\, x_3^2 x_1^2 + (2c_{12} + 4c_{66})\, x_1^2 x_2^2 + \\ & + (4c_{14} + 8c_{56})\, x_1^2 x_2 x_3 + (4c_{25} + 8c_{64})\, x_2^2 x_3 x_1 + (4c_{36} + 8c_{45})\, x_3^2 x_1 x_2 \\ & + 4\,(c_{16} x_1^3 x_2 + c_{15} x_1^3 x_3 + c_{26} x_2^3 x_1 + c_{24} x_2^3 x_3 + c_{35} x_3^3 x_1 + c_{34} x_3^3 x_2\,) \end{aligned} \tag{28}$$

the equivalence (27) allows one to establish the following relations :

$$D_1 = (c_{11})^{-1/4}, \quad D_2 = (c_{22})^{-1/4}, \quad D_3 = (c_{33})^{-1/4} \tag{29}$$

$$c_{23} + 2c_{44} = \sqrt{c_{22}c_{33}}\,, \quad c_{13} + 2c_{55} = \sqrt{c_{11}c_{33}}\,, \quad c_{12} + 2c_{66} = \sqrt{c_{11}c_{22}} \tag{30}$$

$$c_{14} + 2\,c_{56} = c_{25} + 2\,c_{46} = c_{36} + 2\,c_{45} = 0 \tag{31}$$

$$c_{16} = c_{15} = c_{26} = c_{24} = c_{35} = c_{34} = 0 \tag{32}$$

An element of $\hat{\Phi}_4(\mathcal{B})$ thus depends upon 9 intrinsic parameters ($c_{11}$, $c_{22}$, $c_{33}$, $c_{12}$, $c_{23}$, $c_{13}$, $c_{14}$, $c_{25}$, $c_{36}$) and reads :

$$\hat{\Phi}_4(\mathcal{B}) : \qquad \boldsymbol{C} = \begin{bmatrix} c_{11} & c_{12} & c_{13} & c_{14} & & \\ & c_{22} & c_{23} & & c_{25} & \\ & & c_{33} & & & c_{36} \\ & & & \frac{\sqrt{c_{22}c_{33}} - c_{23}}{2} & -\frac{c_{36}}{2} & -\frac{c_{25}}{2} \\ & & & & \frac{\sqrt{c_{11}c_{33}} - c_{13}}{2} & -\frac{c_{14}}{2} \\ & & & & & \frac{\sqrt{c_{11}c_{22}} - c_{12}}{2} \end{bmatrix} \tag{33}$$

The general expressions of $\boldsymbol{C}$ for the class $\hat{\Phi}_2(\mathcal{B})$, as well as the expression of $\underline{\boldsymbol{S}}$ for $\hat{\Gamma}_4(\mathcal{B})$ and $\hat{\Gamma}_2(\mathcal{B})$ can be established in the same way. For instance, for $\hat{\Phi}_2(\mathcal{B})$, $\boldsymbol{C}$ has the same expression (33), but $c_{44}$, $c_{55}$, $c_{66}$ are given by (21) instead of (8), and the relation (29) becomes:

$$D_1 = 1/\sqrt{c_{11}}, \quad D_2 = 1/\sqrt{c_{22}}, \quad D_3 = 1/\sqrt{c_{33}} \tag{34}$$

$\hat{\Gamma}_4(\mathcal{B})$ is found to be defined by the following expression of $\underline{\boldsymbol{S}}$ in which $G_{12}$, $G_{23}$, $G_{13}$ are given by (11) :

$$\hat{\Gamma}_4(\mathcal{B}):\qquad \underline{\underline{S}} = \begin{bmatrix} \frac{1}{E_1} & \frac{-\nu_{12}}{E_1} & \frac{-\nu_{13}}{E_1} & \underline{s}_{14} & & \\ \frac{-\nu_{12}}{E_1} & \frac{1}{E_2} & \frac{-\nu_{23}}{E_2} & & \underline{s}_{25} & \\ \frac{-\nu_{13}}{E_1} & \frac{-\nu_{23}}{E_2} & \frac{1}{E_3} & & & \underline{s}_{36} \\ & & & \frac{1}{G_{23}} & -\underline{s}_{36} & -\underline{s}_{25} \\ & & & & \frac{1}{G_{13}} & -\underline{s}_{14} \\ & & & & & \frac{1}{G_{12}} \end{bmatrix} \tag{35}$$

The surfaces f($\mathbb{C}$) and g($\mathbb{C}$) are always ellipsoidal. The class $\hat{\varphi}(\mathcal{B})$ (respectively $\hat{\gamma}(\mathcal{B})$) represents the tensors $\mathbb{C}$ for which the ellipsoid f($\mathbb{C}$) (respectively g($\mathbb{C}$)) has the principal axes $\mathcal{B}$. Let $\mathbb{C}\in\hat{\varphi}(\mathcal{B})$, and note ($d_1, d_2, d_3$) the semi-diameters of f($\mathbb{C}$). By comparing the (26) like equation of this ellipsoid to that given in the Table (1) for $F_4(\mathbb{C})$, and by using the expression :

$$\begin{aligned}\boldsymbol{\delta}:\mathbb{C}:(\mathbf{x}\otimes\mathbf{x}) = {} & (c_{11}+c_{21}+c_{31})\,x_1^2+(c_{12}+c_{22}+c_{32})\,x_2^2+(c_{13}+c_{23}+c_{33})\,x_3^2+ \\ & 2(c_{14}+c_{24}+c_{34})\,x_2x_3+2(c_{15}+c_{25}+c_{35})\,x_1x_3+2(c_{16}+c_{26}+c_{36})\,x_1x_2\end{aligned} \tag{36}$$

one finds :

$$d_1=\frac{1}{\sqrt{c_{11}+c_{21}+c_{31}}},\quad d_2=\frac{1}{\sqrt{c_{12}+c_{22}+c_{32}}},\quad d_3=\frac{1}{\sqrt{c_{13}+c_{23}+c_{33}}} \tag{37}$$

$$c_{14}+c_{24}+c_{34} = c_{15}+c_{25}+c_{35} = c_{16}+c_{26}+c_{36} = 0 \tag{38}$$

The class $\hat{\gamma}(\mathcal{B})$ is characterized by the relations analogous to (38) satisfied by the coefficients $s_{\alpha\beta}$.

Let us now consider $\mathbb{C}\in\hat{\Phi}_4(\mathcal{B})$ defined by (33), and define the tensor **H** diagonal in $\mathcal{B}$ and given by :

$$\mathbf{H} = \mathrm{diag}\left(\sqrt{c_{11}},\ \sqrt{c_{22}},\ \sqrt{c_{33}}\right) \tag{39}$$

then one finds :

$$\forall\, \mathbf{x};\qquad (\mathbf{x}\otimes\mathbf{x}):\mathbb{C}:(\mathbf{x}\otimes\mathbf{x}) = (\mathbf{x.H.x})^2 \tag{40}$$

Conversely, if a symmetric matrix **H** exists which allows us to write (40), then $F_4(\mathbb{C})$ obviously is ellipsoidal. Therefore we can write:

$$\hat{\Phi}_4 = \{\ \mathbb{C}\ \ /\ \exists\, \mathbf{H};\ \forall\, \mathbf{x};\quad (\mathbf{x}\otimes\mathbf{x}):\mathbb{C}:(\mathbf{x}\otimes\mathbf{x}) = (\mathbf{x.H.x})^2\ \} \tag{41}$$

In the same way :

$$\hat{\Phi}_2 = \{\ \mathbb{C}\ \ /\ \exists\, \mathbf{H};\ \forall\, \mathbf{x};\quad (\mathbf{x}\otimes\mathbf{x}):\mathbb{C}:(\mathbf{x}\otimes\mathbf{x}) = (\mathbf{x.H.x})(\mathbf{x.x})\} \tag{42}$$

$$\hat{\Gamma}_4 = \{\ \mathbb{C}\ \ /\ \exists\, \mathbf{H};\ \forall\, \mathbf{x};\quad (\mathbf{x}\otimes\mathbf{x}):\mathbb{S}:(\mathbf{x}\otimes\mathbf{x}) = (\mathbf{x.H.x})^2\ \} \tag{43}$$

$$\hat{\Gamma}_2 = \{\ \mathbb{C}\ \ /\ \exists\, \mathbf{H};\ \forall\, \mathbf{x};\quad (\mathbf{x}\otimes\mathbf{x}):\mathbb{S}:(\mathbf{x}\otimes\mathbf{x}) = (\mathbf{x.H.x})(\mathbf{x.x})\} \tag{44}$$

It is interesting to notice that the materials of the classes $\hat{\Phi}_4$, $\hat{\Phi}_2$, $\hat{\Gamma}_4$ and $\hat{\Gamma}_2$ in general are not orthotropic, and even can have not any plane of symmetry. As a matter of fact, a plane of symmetry of $\mathbb{C}$ is necessarily a plane of symmetry of its indicator surfaces. If the three semi-diameters of the ellipsoid $F_4(\mathbb{C})$, for instance, are different then its only planes of symmetry are those of the system $\mathcal{B}$, which manifestly don't constitute planes of symmetry for ***C*** given in (33). In consequence, the method of characterization of the material's anisotropy based on the analysis of the number and orientation of its planes of reflective symmetry (Cowin and Mehrabadi 1987, 1995) would not give significant result for these classes of materials.

## 6. Material identification

In this section we study some cases in which two ellipsoidal surfaces allows us to fully determine $\mathbb{C}$.

### 6.1. Intersection of $\hat{\Phi}_4(\mathcal{B})$ and $\hat{\varphi}(\mathcal{B})$

Let us suppose that $F_4(\mathbb{C})$ is ellipsoidal. Since $f(\mathbb{C})$ is always ellipsoidal, the interesting case to consider is when the two ellipsoids have the same principal axes, *i.e.*, when $\mathbb{C} \in \hat{\Phi}_4(\mathcal{B}) \cap \hat{\varphi}(\mathcal{B})$. Then ***C*** will be given by (33), and the condition (38) on $\hat{\varphi}(\mathcal{B})$ implies that $c_{14} = c_{25} = c_{36} = 0$ ; so $\mathbb{C}$ is orthotropic. The following result in this way is obtained :

$$\hat{\Phi}_4(\mathcal{B}) \cap \hat{\varphi}(\mathcal{B}) = \Phi_4(\mathcal{B}) \qquad (45)$$

In the same way, it can be shown that :

$$\hat{\Phi}_2(\mathcal{B}) \cap \hat{\varphi}(\mathcal{B}) = \Phi_2(\mathcal{B}) \qquad (46)$$

The same data c(**n**) can be utilized for defining two surfaces $F_2(\mathbb{C})$ and $F_4(\mathbb{C})$. According that a best ellipsoidal fitting is obtained for $F_4(\mathbb{C})$ or for $F_2(\mathbb{C})$, an element of $\hat{\Phi}_4$ or of $\hat{\Phi}_2$ will be found. The relations (45) and (46) mean that, when fitting $F_4(\mathbb{C})$ or $F_2(\mathbb{C})$ by un ellipsoid, if the principal axes of the ellipsoid are constrained to have the same direction that for $f(\mathbb{C})$, then an orthotropic material will be found. The parameters of $\mathbb{C}$ can be, in these cases, fully determined from the semi-diameters of the two ellipsoids by using (29), (30) and (37) in the first case and (21), (34) and (37) in the second.

Let us consider now the case of materials for which c(**n**) is constant. In this case $\mathbb{C} \in \hat{\Phi}_s$. Since for every system of axes $\mathcal{B}$ we have $\hat{\Phi}_s \subset \hat{\Phi}_4(\mathcal{B})$, one deduces from (45) that $\hat{\Phi}_s \cap \hat{\varphi}(\mathcal{B}) \subset \Phi_4(\mathcal{B})$, and then can establish :

$$\hat{\Phi}_s \cap \hat{\varphi}(\mathcal{B}) = \Phi_s(\mathcal{B}) \qquad (47)$$

An element of $\Phi_s(\mathcal{B})$ is given by (33) in which in which $c_{14} = c_{25} = c_{36} = 0$ and $c_{11} = c_{22} = c_{33}$. If, in addition, $f(\mathbb{C})$ is spherical, then by writing $d_1=d_2=d_3$ in (37) for an element of $\Phi_s(\mathcal{B})$, one finds that $\mathbb{C}$ is isotropic. The relation (47) reduces in this case to the result given be Hé and Curnier (1995) :

$$\hat{\Phi}_s \cap \hat{\varphi}_s = I_s \qquad (48)$$

Equivalent relations to (45), (46) and (47) can be written for $\hat{\Gamma}$ type materials, for instance, for the first relation, $\hat{\Gamma}_4(\mathcal{B}) \cap \hat{\gamma}(\mathcal{B}) = \Gamma_4(\mathcal{B})$.

It is interesting to notice that (47) implies that the elements of $\hat{\Phi}_s$ (and equivalently $\hat{\Gamma}_s$) are all orthotropic. As a matter of fact, for every $\mathbb{C} \in \hat{\Phi}_s$, if $\mathcal{B}$ denotes a system of principal axes of $f(\mathbb{C})$, then $\mathbb{C} \in \hat{\varphi}(\mathcal{B})$, and then according to (47), $\mathbb{C} \in \Phi_s(\mathcal{B})$ and so $\mathbb{C}$ is orthotropic:

*If the elastic coefficient* c(**n**) *(or the Young's modulus* E(**n**)*) is constant in all directions, then the material is orthotropic.*

### 6.2. Intersection of $\hat{\Phi}_4$ and $\hat{\Gamma}_4$

We consider in this section the case of materials for which both indicator surfaces $F_4(\mathbb{C})$ and $G_4(\mathbb{C})$ are ellipsoidal. We first consider the case in which these ellipsoids have a common set of principal axes $\mathcal{B}$. Then the more general case will be considered in which this condition is not supposed *a priori*.

#### 6.2.1. Intersection of $\hat{\Phi}_4(\mathcal{B})$ and $\hat{\Gamma}_4(\mathcal{B})$

If $\mathbb{C} \in \hat{\Phi}_4(\mathcal{B}) \cap \hat{\Gamma}_4(\mathcal{B})$, then the planes of the coordinate system $\mathcal{B}$ constitute planes of reflexive symmetry for the surfaces $F_4(\mathbb{C})$ and $G_4(\mathbb{C})$, or also for the scalar functions $(\mathbf{n}\otimes\mathbf{n}):\mathbb{C}:(\mathbf{n}\otimes\mathbf{n})$ and $(\mathbf{n}\otimes\mathbf{n}):\mathbb{S}:(\mathbf{n}\otimes\mathbf{n})$. This property for the function $(\mathbf{n}\otimes\mathbf{n}):\mathbb{C}:(\mathbf{n}\otimes\mathbf{n})$ is equivalent to the relations (30) and (31) and also to the expression of the matrix $\boldsymbol{C}$ given by (49) and (50). The same property for $(\mathbf{n}\otimes\mathbf{n}):\mathbb{S}:(\mathbf{n}\otimes\mathbf{n})$ is equivalent to the $\underline{\boldsymbol{S}}$ given by (49) and (51):

$$\boldsymbol{C} = \begin{bmatrix} \mathbf{M} & \mathbf{D} \\ \mathbf{D} & \frac{1}{2}\mathbf{N} \end{bmatrix}, \quad \underline{\boldsymbol{S}} = \begin{bmatrix} \mathbf{M}' & 2\mathbf{D}' \\ 2\mathbf{D}' & 2\mathbf{N}' \end{bmatrix} \tag{49}$$

with :

$$\mathbf{M}=\begin{bmatrix} c_{11} & c_{12} & c_{13} \\ & c_{22} & c_{23} \\ & & c_{33} \end{bmatrix}, \quad \mathbf{N} = \begin{bmatrix} 2c_{44} & -c_{36} & -c_{25} \\ & 2c_{55} & -c_{14} \\ & & 2c_{66} \end{bmatrix}, \quad \mathbf{D}=\begin{bmatrix} c_{14} & & \\ & c_{25} & \\ & & c_{36} \end{bmatrix} \tag{50}$$

$$\mathbf{M}'=\begin{bmatrix} \underline{s}_{11} & \underline{s}_{12} & \underline{s}_{13} \\ & \underline{s}_{22} & \underline{s}_{23} \\ & & \underline{s}_{33} \end{bmatrix}, \quad \mathbf{N}'= \frac{1}{2}\begin{bmatrix} \underline{s}_{44} & -\underline{s}_{36} & -\underline{s}_{25} \\ & \underline{s}_{55} & -\underline{s}_{14} \\ & & \underline{s}_{66} \end{bmatrix}, \quad \mathbf{D}'=\frac{1}{2}\begin{bmatrix} \underline{s}_{14} & & \\ & \underline{s}_{25} & \\ & & \underline{s}_{36} \end{bmatrix} \tag{51}$$

The equation $\mathbb{C}:\mathbb{S} = \mathbb{I}$, or $c_{\alpha\beta}\, \underline{s}_{\beta\gamma} = \delta_{\alpha\beta}$, implies the following matrix equations in which $\mathbf{I}$ represents the 3×3 unit matrix :

$$\mathbf{MM}' + 2\mathbf{DD}' = \mathbf{I} \tag{52}$$
$$\mathbf{NN}' + 2\mathbf{DD}' = \mathbf{I} \tag{53}$$
$$\mathbf{MD}' + \mathbf{DN}' = \mathbf{0} \tag{54}$$
$$\mathbf{DM}' + \mathbf{ND}' = \mathbf{0} \tag{55}$$

Symmetry and positive definite properties of $\mathbb{C}$ and $\mathbb{S}$ imply that $\mathbf{M}$, $\mathbf{M}'$, $\mathbf{N}$ and $\mathbf{N}'$ are symmetric and positive definite.

In addition to these relations, the equation (8) and (10) must be satisfied for $F_4(\mathbb{C})$ and $G_4(\mathbb{C})$ to be ellipsoidal.

The equations (50) to (55) which express the symmetry of $(\mathbf{n}\otimes\mathbf{n}):\mathbb{C}:(\mathbf{n}\otimes\mathbf{n})$ and $(\mathbf{n}\otimes\mathbf{n}):\mathbb{S}:(\mathbf{n}\otimes\mathbf{n})$ with respect to the planes of the coordinate system have been studied in the Appendix 2. It has been shown that they can have three different types of solutions and we show here that only one of these types is compatible with (8) and (10).

The first type (Case 1.1) is characterized by a parameter $\eta_1 \neq 0$ and by, in particular, the following relations between the parameters (equations (11) and (10) in Appendix 2):

$\underline{s}_{44} = (1-2\eta_1)/2c_{44}$ , $\underline{s}_{22} = c_{33}/[c_{22}c_{33}-(c_{23})^2]$, $\underline{s}_{33} = c_{22}/[c_{22}c_{33}-(c_{23})^2]$, $\underline{s}_{23} = -c_{23}/[c_{22}c_{33}-(c_{23})^2]$

Substituting by these relations in the first equation of (10), one finds :

$$\frac{1-2\eta_1}{2\,c_{44}} = \frac{\sqrt{c_{22}\;c_{33}}}{c_{22}\;c_{33}-c_{23}^2} + \frac{c_{23}}{c_{22}\;c_{33}-c_{23}^2} = \frac{1}{\sqrt{c_{22}\;c_{33}}-c_{23}}$$

Since this equation compared to (8) implies $\eta_1 = 0$, the Case 1 is incompatible with (8) and (10).

The second type of solutions (Case 2.1) is characterized by a parameter $\eta<0$ and a parameter $\kappa$ satisfying :

$$(1-4\eta)^2\kappa^2 + (2-9\eta)\eta\kappa + \eta^2 = 0 \qquad (56)$$

In this case we have in particular the following relation between the parameters (deduced from (34) and (36) in Appendix 2):

$$2\,c_{44} + c_{33} = \beta^2(1-\kappa)\sqrt{c_{22}c_{33}}$$

where $\beta = -\eta/[\kappa(1-4\eta)]$. This equation compared to (8) leads to $\beta^2(1-\kappa) = 1$ and then to $\eta^2(1-\kappa)/[\kappa(1-4\eta)]^2 = 1$. Substituting in this equation for $\kappa^2(1-4\eta)^2$ by (56), one finds $\eta[\kappa(1-5\eta)+\eta] = 0$. Since $\eta = 0$ is not compatible with Case 2.1, we deduce $\kappa = -\eta/(1-5\eta)$. By substituting by this expression of $\kappa$ in (56) one finds $\eta^3(1-4\eta) = 0$ which is also incompatible with the condition $\eta<0$ characterizing the Case 2.1. Thus, this case is also incompatible with (8).

Therefore the only type of solution of (50) to (55) compatible with (8) and (9) is the last type described in Appendix 2 (Case 2.2) in which **D** = **D'** = **0**.

In conclusion for $\mathbb{C} \in \hat{\Phi}_4(\mathcal{B}) \cap \hat{\Gamma}_4(\mathcal{B})$, the matrices ***C*** and ***S̲*** are given in the coordinate system $\mathcal{B}$ by (49) with **D** = **D'** = **0.** This means that $\mathbb{C} \in \hat{\Phi}_4(\mathcal{B}) \cap \hat{\Gamma}_4(\mathcal{B})$ is orthotropic and that its planes of orthotropy are given by $\mathcal{B}$. This is a main result of the present paper. It allows us write :

$$\hat{\Phi}_4(\mathcal{B})\cap\hat{\Gamma}_4(\mathcal{B}) = \Phi_4(\mathcal{B})\cap\Gamma_4(\mathcal{B}) \qquad (57)$$

and, by using (12) :

$$\hat{\Phi}_4(\mathcal{B})\cap\hat{\Gamma}_4(\mathcal{B}) = \Psi(\mathcal{B}) \qquad (58)$$

In this case, the matrices ***C*** and ***S̲*** respectively are given by (13) and (14). The parameters $c_{11}$, $c_{22}$, $c_{33}$ can be deduced from the semi-diameters of $F_4(\mathbb{C})$ (relations (29)) and the semi-diameters $\{\sqrt[4]{E_1}, \sqrt[4]{E_2}, \sqrt[4]{E_3}\}$ of $G_4(\mathbb{C})$, and $\eta$ and $\nu$ are deduced from (15). The tensor $\mathbb{C}$ in this way is fully determined.

### 6.2.2. Intersections of $\hat{\Phi}_s$ and $\hat{\Gamma}_s$

We first study $\hat{\Phi}_s\cap\hat{\Gamma}_4(\mathcal{B})$. Since $\hat{\Phi}_s\subset\hat{\Phi}_4(\mathcal{B})$, (58) implies that $\hat{\Phi}_s\cap\hat{\Gamma}_4(\mathcal{B})\subset\Psi(\mathcal{B})$. Then if $\mathbb{C}\in\hat{\Phi}_s\cap\hat{\Gamma}_4(\mathcal{B})$, ***C*** is given by (13) and the relations (29) lead to $c_{11} = c_{22} = c_{33}$. This means that $\mathbb{C}$ is isotropic. We deduce that $\hat{\Phi}_s\cap\hat{\Gamma}_4(\mathcal{B})\subset I_s$ and then show that:

$$\hat{\Phi}_s\cap\hat{\Gamma}_4(\mathcal{B}) = I_s \qquad (59)$$

By using the definition (24) of $\hat{\Gamma}_4$, the equality (59) can be extended to :

$$\hat{\Phi}_s \cap \hat{\Gamma}_4 = I_s \qquad (60)$$

Since $\hat{\Gamma}_s \subset \hat{\Gamma}_4$, we deduce from (60) that $\hat{\Phi}_s \cap \hat{\Gamma}_s \subset I_s$ and since conversely $I_s \subset \hat{\Phi}_s$ and $I_s \subset \hat{\Gamma}_s$, then:

$$\hat{\Phi}_s \cap \hat{\Gamma}_s = I_s \qquad (61)$$

This result means that : *if* E(**n**) *and* c(**n**) *both are constant in all the directions, then* $\mathbb{C}$ *is isotropic*.

The coefficients E and ν corresponding to this isotropic elasticity tensor can be deduced from the radii of the spheres $F_4(\mathbb{C})$ and $G_4(\mathbb{C})$, noted respectively R and ρ, by :

$$\rho^4 = E \qquad , \qquad R^4 = \frac{(1+\nu)(1-2\nu)}{(1-\nu)E} \qquad (62)$$

### 6.2.3. Intersection of $\hat{\Phi}_4$ and $\hat{\Gamma}_4$

Let us now suppose that $\mathbb{C} \in \hat{\Phi}_4(\mathcal{B}) \cap \hat{\Gamma}_4$. In this case, $\mathcal{B}$ defines a system of principal axes for the ellipsoid $F_4(\mathbb{C})$, but, *a priori*, not necessarily for $G_4(\mathbb{C})$. Since $\mathbb{C} \in \hat{\Phi}_4(\mathcal{B})$, the matrix ***C*** has the expression (33) in the coordinate system $\mathcal{B}$. Consider now the tensor **Q**, diagonal in the basis $\mathcal{B}$, and given by :

$$\mathbf{Q} = \mathrm{diag}\left(\frac{1}{\sqrt[4]{c_{11}}},\ \frac{1}{\sqrt[4]{c_{22}}},\ \frac{1}{\sqrt[4]{c_{33}}}\right) \qquad (63)$$

Let the tensor $\mathbb{C}'$ be deduced from $\mathbb{C}$ by the following *transformation* (Pouya 2000, Pouya and Zaoui 2005) :

$$C'_{ijkl} = Q_{im}\, Q_{jn}\, Q_{kp}\, Q_{lq}\, C_{mnpq} \qquad (64)$$

It can be verified that $\mathbb{C}'$ has the symmetries of an elasticity tensor and is positive definite. So it represents a new elasticity tensor. The matrix ***C'*** has the expression (33) in the coordinate system $\mathcal{B}$ with, in particular :

$$c'_{11} = 1,\ c'_{12} = c_{12}/\sqrt{c_{11}c_{22}}\ ,\ c'_{14} = c_{14}/\sqrt[4]{c_{11}^{\,2}c_{22}c_{33}}\ ,\ c'_{44} = c_{44}/\sqrt{c_{22}c_{33}}\ ,\ c'_{45} = c_{45}/\sqrt[4]{c_{11}^{\,2}c_{22}c_{33}} \qquad (65)$$

The other elements of ***C'*** are deduced from (65) by index permutation. The relations (30) and (31) are well satisfied; for instance :

$$c'_{44} = \frac{\sqrt{c'_{22}\,c'_{33}} - c'_{23}}{2} \qquad , \qquad c'_{45} = -\,c'_{36}/2$$

Since $c'_{11} = c'_{22} = c'_{33} = 1$ then $\mathbb{C}' \in \hat{\Phi}_s$. Denoting $\mathbb{S}' = \mathbb{C}'^{-1}$ and $\mathbf{P} = \mathbf{Q}^{-1}$, we deduce from (64) :

$$S'_{ijkl} = P_{im}\, P_{jn}\, P_{kp}\, P_{lq}\, S_{mnpq} \qquad (66)$$

Since it is assumed that $\mathbb{C} \in \hat{\Gamma}_4$, according to (43), a reversible tensor **T** exists which allows us to write :

$$\forall\ \mathbf{x}\ ; \qquad (\mathbf{x}\otimes\mathbf{x}):\mathbb{S}:(\mathbf{x}\otimes\mathbf{x}) = (\mathbf{x}.\mathbf{T}.\mathbf{x})^2 \qquad (67)$$

The relation (66) then implies that :

$$\forall\ \mathbf{x}\ ; \qquad (\mathbf{x}\otimes\mathbf{x}):\mathbb{S}':(\mathbf{x}\otimes\mathbf{x}) = (\mathbf{x}.\mathbf{T}'.\mathbf{x})^2 \qquad (68)$$

with:

$$\mathbf{T}' = \mathbf{PTP}$$

According to (43), this means that $\mathbb{C}'\in\hat{\Gamma}_4$. Since already $\mathbb{C}'\in\hat{\Phi}_s$, the relation (60) implies that $\mathbb{C}'\in I_s$. Then writing (64) as :

$$C_{ijkl} = P_{im}\, P_{jn}\, P_{kp}\, P_{lq}\, C'_{mnpq}$$

and comparing to (16), we deduce that $\mathbb{C}\in\Psi(\mathcal{B})$. This implies that $\hat{\Phi}_4(\mathcal{B})\cap\hat{\Gamma}_4 \subseteq \Psi_4(\mathcal{B})$. Since conversely $\Psi(\mathcal{B}) \subset\hat{\Phi}_4(\mathcal{B})$ and $\Psi(\mathcal{B}) \subset\hat{\Gamma}_4$ , we obtain :

$$\hat{\Phi}_4(\mathcal{B})\cap\hat{\Gamma}_4 = \Psi(\mathcal{B}) \tag{69}$$

By using the definitions (19) and (24) of $\Psi$ and of $\hat{\Phi}_4$, the result (69) allow us to write :

$$\hat{\Phi}_4\cap\hat{\Gamma}_4 = \Big(\bigcup_{\mathcal{B}} \hat{\Phi}_4(\mathcal{B})\Big)\cap\hat{\Gamma}_4 = \bigcup_{\mathcal{B}} \Big(\hat{\Phi}_4(\mathcal{B})\cap\hat{\Gamma}_4\Big) = \bigcup_{\mathcal{B}} \Psi(\mathcal{B}) = \Psi$$

and finally:

$$\hat{\Phi}_4 \cap\hat{\Gamma}_4 = \Psi \tag{70}$$

This equation means that *if the surfaces* $F_4(\mathbb{C})$ *and* $G_4(\mathbb{C})$ *both are ellipsoidal, then they have necessarily the same (or a common system of) principal axes*. Moreover, the material is orthotropic and the common system of principal axes of the two ellipsoids defines also the planes of orthotropy.

In consequence, in phenomenological modelling of materials elasticity, when the values of E(**n**) and c(**n**) are given and one searches for an ellipsoidal approximation of the both surfaces $F_4(\mathbb{C})$ and $G_4(\mathbb{C})$, then the two ellipsoids fitting these surfaces must be constrained to have the same directions of principal axes. Only under this condition all the parameters of $\mathbb{C}$, which will be defined in this case by the expressions (13) or (14), can be determined. They will be deduced from the semi-diameters of the two ellipsoids and the relations (15).

## 7. Discussion and conclusions

In this paper, we studied some cases of ellipsoidal anisotropy and the method and conditions of the elasticity tensor determination from the ellipsoidal fitting parameters.

As mentioned here above, the solution of many basic problems of elasticity for $\Psi$ type materials can be deduced by a *linear transformation* from the solutions known for isotropic materials (Pouya 2000, Pouya and Zaoui 2005). It should be emphasized that some problems which can be solved in this way for $\Psi$ type materials, such as the Green function of a half-space (Pouya and Zaoui 2005), can not be approached by other methods such as the Stroh formalism (Stroh 1958, Ting 1996) since they don’t present any plane of symmetry. Our on going investigations show that some of these closed form solutions, for instance the Green function solution for infinite space, can be established for the family $\hat{\Phi}_4$. This increases the interest of this type of anisotropy. The class $\hat{\Phi}_2$ recovers, as mentioned above, a large family of models already used for damaged or heterogeneous materials which are based on a second order damage tensor (Kachanov 1992). Ellipsoidal models could also be used for approximation of more complex damage models based on fourth-order tensors (Zheng 1997). It can be noticed that in spite of more interesting properties of $\hat{\Phi}_4$ and $\hat{\Phi}_2$, these are the classes $\hat{\Gamma}_4$ and $\hat{\Gamma}_2$ which have drawn more attention in the literature. For instance, $\Gamma_4$ and $\Gamma_2$ are the only families mentioned by Lekhnitskii (1963) when quoting Saint Venant’s work.

In conclusion, the concept of ellipsoidal anisotropy, when it fits well the materials data, seems an attractive guideline for phenomenological modelling of the amorphous or damaged materials elasticity. As a way to define anisotropic models with reduced number of parameters, it constitutes an alternative to the method based on the number and orientation of planes of reflective symmetry (Cowin and Mehrabadi 1987, 1995). The two methods recover different families of materials. Besides providing potential tools for modelling the materials anisotropy, ellipsoidal anisotropy classes, which are an extension of the families introduced by Saint Venant, present interesting theoretical features concerning the resolution of elastic body problems.

## Appendix 1

**The indicator surface of the Young's modulus**

The indicator surface of the Young's modulus is defined by the equation (see the main text) :

$$[(\mathbf{x}\otimes\mathbf{x}):\mathbb{S}:(\mathbf{x}\otimes\mathbf{x})]^2 = (\mathbf{x}.\mathbf{x})^3 \qquad (1)$$

Let us suppose that this surface is an ellipsoid. The equation of an ellipsoid can be written as :

$$\mathbf{x}.\mathbf{H}.\mathbf{x} = 1 \qquad (2)$$

where **H** is a second rank symmetric and positive definite tensor. Then, we can write the following equivalence :

$$\forall\ \mathbf{x}\ ;\quad [(\mathbf{x}\otimes\mathbf{x}):\mathbb{S}:(\mathbf{x}\otimes\mathbf{x})]^2 = (\mathbf{x}.\mathbf{x})^3 \ \Leftrightarrow\ \mathbf{x}.\mathbf{H}.\mathbf{x} = 1 \qquad (3)$$

For every vector **x**, if we note $\rho = (\mathbf{x}.\mathbf{H}.\mathbf{x})^{1/2}$, et $\mathbf{x}' = \mathbf{x}/\rho$, we find :

$$\mathbf{x}'.\mathbf{H}.\mathbf{x}' = 1 \qquad (4)$$

Then, according to (3), we have $[(\mathbf{x}'\otimes\mathbf{x}'):\mathbb{S}:(\mathbf{x}'\otimes\mathbf{x}')]^2 = (\mathbf{x}'.\mathbf{x}')^3$, and by taking account of (4), $[(\mathbf{x}'\otimes\mathbf{x}'):\mathbb{S}:(\mathbf{x}'\otimes\mathbf{x}')]^2 = (\mathbf{x}'.\mathbf{x}')^3(\mathbf{x}'.\mathbf{H}.\mathbf{x}')$. Multiplying the two sides of this equality by $\rho^8$, we find: $[(\mathbf{x}\otimes\mathbf{x}):\mathbb{S}:(\mathbf{x}\otimes\mathbf{x})]^2 = (\mathbf{x}.\mathbf{x})^3(\mathbf{x}.\mathbf{H}.\mathbf{x})$. Therefore :

$$\forall\ \mathbf{x}\ ;\qquad [(\mathbf{x}\otimes\mathbf{x}):\mathbb{S}:(\mathbf{x}\otimes\mathbf{x})]^2 = (\mathbf{x}.\mathbf{x})^3(\mathbf{x}.\mathbf{H}.\mathbf{x}) \qquad (5)$$

If we take $\mathbf{x} = (x, 1, 0)$, the expression $(\mathbf{x}\otimes\mathbf{x}):\mathbb{S}:(\mathbf{x}\otimes\mathbf{x})$ will become a fourth order polynomial in the scalar variable $x$. This polynomial hasn't any real root, because if $(\mathbf{x}\otimes\mathbf{x}):\mathbb{S}:(\mathbf{x}\otimes\mathbf{x})=0$, owing to positive definite assumption for $\mathbb{S}$, we must have $\mathbf{x}\otimes\mathbf{x} = 0$, and then $\mathbf{x}=0$, and this is incompatible with $\mathbf{x} = (x, 1, 0)$. Therefore, this forth order polynomial in $x$ can be decomposed in the product of two irreducible second order polynomials :

$$(\mathbf{x}\otimes\mathbf{x}):\mathbb{S}:(\mathbf{x}\otimes\mathbf{x}) = P_1(x)\, P_2(x) \qquad (6)$$

The equation (5) implies then :

$$[P_1(x)]^2\, [P_2(x)]^2 = (x^2+ 1)^3\, (H_{11}\, x^2 + 2H_{12}\, x + H_{22}) \qquad (7)$$

Sine all the polynomials in the two sides of this equality are irreducible, $P_1(x)$ and $P_2(x)$ must be proportional to $x^2+1$ or to $H_{11}\, x^2 + 2H_{12}\, x + H_{22}$. Then, since the polynomials appear in the left side at pair power and, in the right, at odd power, on can establish that (7) is possible only if $H_{11}\, x^2 + 2H_{12}\, x + H_{22}$ is proportional to $x^2+1$. This means that $H_{11} = H_{22}$ et $H_{12} = 0$. In the same way we can establish that $H_{11}= H_{33}$, $H_{13} = 0$ and $H_{23} = 0$, and globally that **H** is proportional to the unit tensor. Therefore, the surface defined by the equation (2) is a sphere.

The indicating surface of the Young's modulus thus can not be an ellipsoid different from a sphere.

## Appendix 2

**Conditions of three orthogonal planes of symmetry for** c(**n**) **and** E(**n**)

The conditions of symmetry of the scalar functions $(\mathbf{n}\otimes\mathbf{n}):\mathbb{C}:(\mathbf{n}\otimes\mathbf{n})$ and $(\mathbf{n}\otimes\mathbf{n}):\mathbb{S}:(\mathbf{n}\otimes\mathbf{n})$ with respect to the planes of the coordinate system has been considered in the main text. They are equivalent to suppose that the following matrices :

$$\mathbf{A} = \begin{bmatrix} a_1 & c_3 & c_2 \\ c_3 & a_2 & c_1 \\ c_2 & c_1 & a_3 \end{bmatrix}, \qquad \mathbf{B} = \begin{bmatrix} b_1 & -d_3 & -d_2 \\ -d_3 & b_2 & -d_1 \\ -d_2 & -d_1 & b_3 \end{bmatrix}, \qquad \mathbf{D} = \begin{bmatrix} d_1 & & \\ & d_2 & \\ & & d_3 \end{bmatrix}$$

$$\mathbf{A'} = \begin{bmatrix} a'_1 & c'_3 & c'_2 \\ c'_3 & a'_2 & c'_1 \\ c'_2 & c'_1 & a'_3 \end{bmatrix}, \qquad \mathbf{B'} = \begin{bmatrix} b'_1 & -d'_3 & -d'_2 \\ -d'_3 & b'_2 & -d'_1 \\ -d'_2 & -d'_1 & b'_3 \end{bmatrix}, \quad \mathbf{D'} = \begin{bmatrix} d'_1 & & \\ & d'_2 & \\ & & d'_3 \end{bmatrix}$$

whit **A**, **A'**, **B** and **B'** positive-definite, satisfy the following equations:

$$\mathbf{AA'} = \mathbf{I} - 2\,\mathbf{DD'} \qquad (1)$$

$$\mathbf{BB'} = \mathbf{I} - 2\,\mathbf{DD'} \qquad (2)$$

$$\mathbf{AD'} + \mathbf{DB'} = 0 \qquad (3)$$

$$\mathbf{DA'} + \mathbf{BD'} = 0 \qquad (4)$$

where **I** is the 3×3 unit matrix. These matrix equations lead to a system of equations on ($a_i$, $b_i$, $c_i$, $d_i$, $a'_i$, $b'_i$, $c'_i$, $d'_i$), with (i=1,2,3), which is invariant for permutation of the indexes {1,2,3}. This property of *index permutation invariance* allows us an extension of some results. Let us denote :

$$\eta_1 = d_1 d'_1, \qquad \eta_2 = d_2 d'_2, \qquad \eta_3 = d_3 d'_3 \qquad (5)$$

We can write :

$$\mathbf{DD'} = \mathbf{D'D} = \begin{bmatrix} \eta_1 & & \\ & \eta_2 & \\ & & \eta_3 \end{bmatrix} \qquad (6)$$

Multiplying the two sides of (4) at right by **D** one finds : **BD'D** = **-DA'D**, and deduces that **BD'D** is symmetric. This symmetry implies the following relations :

$$\eta_1\, d_3 = \eta_2\, d_3\,, \qquad \eta_2\, d_1 = \eta_3\, d_1\,, \qquad \eta_3\, d_2 = \eta_1\, d_2 \qquad (7)$$

Multiplying the two sides of $\eta_1 d_3 = \eta_2 d_3$ by $d'_3$ and by using (5), one finds $\eta_1\eta_3 = \eta_2\eta_3$. Index permutation gives:

$$\eta_1\eta_2 = \eta_2\eta_3 = \eta_3\eta_1 \qquad (8)$$

This equations implies that either $\eta_1 = \eta_2 = \eta_3$ or two of $\eta_i$ are equal to zero and the third one is different. We study in the following the general structure of the equations in these two cases. Because of *index permutation invariance*, the study of the case in which only one $\eta_i$ is different from zero (Case 1) can be restricted to the study of the sub-case $\eta_1 \neq 0$, and $\eta_2 = \eta_3 = 0$ (Case 1.1). In the other case (Case 2), the sub-cases $\eta_1 = \eta_2 = \eta_3 \neq 0$ (Case 2.1) and $\eta_1 = \eta_2 = \eta_3 = 0$ (Case 2.2) will be distinguished.

### 1.1. Case 1.1

In this case $\eta_1 \neq 0$ and $\eta_2 = \eta_3 = 0$. All the equations and assumptions are invariant for index permutation {2,3}. The equations (7) imply : $d_2 = d_3 = 0$. Multiplying the both sides of (3) at left by **D**, one finds : **AD'D** = **-DB'D** = 0. Then, writing $(\mathbf{A.D'.D})_{21} = -(\mathbf{D.B'.D})_{21}$ one finds : $c_3\eta_1 = d_2 d'_3 d_1 = 0$, and it implies $c_3 = 0$, and by index permutation, $c_2 = 0$. Therefore the matrices **A**, **B** and **D** have the following expressions :

$$\mathbf{A} = \begin{bmatrix} a_1 & 0 & 0 \\ 0 & a_2 & c_1 \\ 0 & c_1 & a_3 \end{bmatrix}, \ \mathbf{B} = \begin{bmatrix} b_1 & 0 & 0 \\ 0 & b_2 & -d_1 \\ 0 & -d_1 & b_3 \end{bmatrix}, \ \mathbf{D} = \begin{bmatrix} d_1 & & \\ & 0 & \\ & & 0 \end{bmatrix}, \ \mathbf{DD'} = \begin{bmatrix} \eta_1 & & \\ & 0 & \\ & & 0 \end{bmatrix} \qquad (9)$$

**A'** et **B'** can be deduced from (9) and (1) and (2). The matrix **D'** can also be determined since its elements are given by the non diagonal terms of **B'**. One will find that these matrices have the same form as **A**, **B** and **D** with :

$a'_1 = (1-2\eta_1)/a_1$, $\quad a'_2 = a_3/(a_2a_3-c_1^2)$, $\quad a'_3 = a_2/(a_2a_3-c_1^2)$, $\quad c'_1 = -c_1/(a_2a_3-c_1^2)$ (10)

$b'_1 = (1-2\eta_1)/b_1$, $\quad b'_2 = b_3/(b_2b_3-d_1^2)$, $\quad b'_3 = b_2/(b_2b_3-d_1^2)$, $\quad d'_1 = -d_1/(b_2b_3- d_1^2)$ (11)

The parameters defining **A**, **B** and **D** are not all independent. As a matter of fact, (2) implies $(\mathbf{BB'})_{22} = 1$, and then : $b_2b'_2+d'_1d_1 = 1$. By substituting in this equation for $b'_2$ by (11) and using $d'_1d_1 = \eta_1$, one finds :

$$d_1^2 = -\eta_1b_2b_3/(1-\eta_1) \qquad (12)$$

The equation (3) implies $(\mathbf{AD'}+\mathbf{DB'})_{11} = 0$, and then: $a_1d'_1+d_1b'_1 = 0$. Multiplication of the two sides by $d_1$ yields: $a_1\eta_1+d_1^2b'_1 = 0$, and substitution $b'_1$ by (11) and for $d_1^2$ by (12) gives:

$$\frac{a_1b_1}{b_2b_3} = \frac{1-2\eta_1}{1-\eta_1} \qquad (13)$$

In conclusion, in this case all the parameters of **A**, **B**, **D**, **A'**, **B'**, **D'**, except for the sign of $d_1$ and $d'_1$, are deduced from ($a_1$, $a_2$, $a_3$, $c_1$, $b_1$, $b_2$, $b_3$) by the equations (13), (12), (10) and (11).

**2. Case 2**

In this case:

$$\eta_1 = \eta_2 = \eta_3 = \eta \qquad (14)$$

and all the equations are invariant for index permutation {1,2,3}. The relation (6) can be written as :

$$\mathbf{D'D} = \eta\,\mathbf{I} \qquad (15)$$

By multiplying the two sides of (3) at right by **D**, we find :

$$\mathbf{DB'D} = -\eta\mathbf{A} \qquad (16)$$

and we deduce $det\mathbf{B'}\,(det\mathbf{D})^2 = -\eta^3\, det\mathbf{A}$. Since **A** and **B'** are positive-definite this relation implies:

$$\eta \le 0 \qquad (17)$$

Besides, taking account of (15), the relation (2) reads: $\mathbf{BB'} = (1-2\eta)\,\mathbf{I}$, and implies $(\mathbf{BB'})_{11} = (1-2\eta)$, or also $b_1b'_1 + d_3d'_3 + d_2d'_2 = (1-2\eta)$. Substituting in this equation for $d_3d'_3$ and $d_2d'_2$ by $\eta$, one finds $b_1b'_1 = 1-4\eta$. Index permutation leads to :

$$b_1b'_1 = b_2b'_2 = b_3b'_3 = 1-4\eta \qquad (18)$$

The elements of $\mathbf{B}^{-1}$ can be explicitly expressed in terms of **B**. In particular we have :

$$(\mathbf{B}^{-1})_{11} = (b_2\, b_3 - d_1^2)/\delta, \qquad (\mathbf{B}^{-1})_{13} = (d_2\, d_3 + b_1d_1)/\delta \qquad (19)$$

where :

$$\delta = det\mathbf{B} = b_1b_2b_3 - (b_1d_1^2 + b_2d_2^2 + b_3d_3^2) - 2\, d_1d_2d_3 \qquad (20)$$

Using (19) and $\mathbf{B'} = (1-2\eta)\,\mathbf{B}^{-1}$, we find :

$$b'_1 = (1-2\eta)\,(b_2\, b_3 - d_1^2)/\delta \qquad (21)$$

$$d'_1 = -(1-2\eta)\,(d_2\, d_3 + b_1d_1)/\delta \qquad (22)$$

Multiplying the both sides of (21) by $b_1$, one finds $b_1b'_1 = (1-2\eta)\,(b_1\, b_2\, b_3 - b_1\, d_1^2)/\delta$ and then using (18) one finds :

$$b_1\, d_1^2 = b_1\, b_2\, b_3 - \delta(1-4\eta)/(1-2\eta) \qquad (23)$$

We note :

$$b = b_1b_2b_3 \qquad (24)$$

Index permutation invariance in (23) implies $b_1\, d_1^2 = b_2\, d_2^2 = b_3\, d_3^2$. We can define $\kappa$ by writing :

$$b_1\, d_1^2 = b_2\, d_2^2 = b_3\, d_3^2 = b\kappa \qquad (25)$$

Multiplying the two sides of (22) by $d_1$ and using $d_1d'_1 = \eta$ and $b_1d_1^2 = b\kappa$, one finds :

$$d_1d_2d_3 = -b\kappa - \delta\eta/(1-2\eta) \qquad (26)$$

Substituting in (20) by (23) and (24), one finds :

$$\delta = b(1-3\kappa) - 2\, d_1d_2d_3 \qquad (27)$$

Elimination of $\delta$ and $d_1d_2d_3$ between (27) and (26) leads to :

$$\delta = b(1-\kappa)(1-2\eta)/(1-4\eta) \qquad (28)$$

$$d_1d_2d_3 = -b\,[\eta+\kappa(1-5\eta)]/(1-4\eta) \qquad (29)$$

The equations (25) imply $b^3\kappa^3 = b_1b_2b_3\, d_1^2d_2^2d_3^2 = b(d_1d_2d_3)^2$. Substituting in this equation for $d_1d_2d_3$ by (29), one finds for $\kappa$ the equation: $(1-\kappa)[(1-4\eta)^2\kappa^2 + (2-9\eta)\eta\kappa + \eta^2\,]=0$. The assumption that **B** is positive-definite implies that $\delta > 0$ and also $b_1 > 0$, $b_2 > 0$, $b_3 > 0$, and thus $b>0$. Then, since $\eta \le 0$, (28) implies $1-\kappa > 0$, the equation on $\kappa$ reduces to:

$$(1-4\eta)^2\kappa^2 + (2-9\eta)\eta\kappa + \eta^2 = 0 \qquad (30)$$

Now, we will consider the two sub-cases $\eta \neq 0$ (Case 2.1) and $\eta=0$ (Case 2.2).

### 2.1. Case 2.1

In this case $\eta < 0$. We deduce from (16) that $\eta\mathbf{A}_{11} = -(\mathbf{DB'D})_{11}$, and then $a_1 = -d_1^2 b'_1/\eta$. By multiplying the two sides by $b_1^2$ and by taking account of (18) and (25), one finds $a_1b_1^2 = -b\kappa\,(1-4\eta)/\eta$. Index permutation invariance then gives :

$$a_1b_1^2 = a_2b_2^2 = a_3b_3^2 = -b\kappa(1-4\eta)/\eta \qquad (31)$$

This equation allows us to write $-[b\kappa(1-4\eta)/\eta]^3 = a_1b_1^2a_2b_2^2a_3b_3^2 = a_1a_2a_3b^2$ and to deduce :

$$b = \beta^3\, a_1a_2a_3 \ , \qquad \beta = -\eta/[\kappa(1-4\eta)] \qquad (32)$$

From (16) we deduce also $-\eta\mathbf{A}_{23} = (\mathbf{DB'D})_{23}$, and so $\eta c_1 = d_2d_3d'_1$. By multiplying the two sides of this equality by $d_1^2$ and simplifying by $\eta$, one finds $c_1d_1^2 = d_1d_2d_3$. By index permutation and by using (29) one finds:

$$c_1d_1^2 = c_2d_2^2 = c_3d_3^2 = -b\,[\eta+\kappa(1-5\eta)]/(1-4\eta) \qquad (33)$$

The equation (30) has two negative roots, compatible with the condition $\delta > 0$ in (28). Now, we will show that if $(a_1, a_2, a_3, \eta)$ are given and if one of the roots $\kappa$ of (30) is chosen, then all the parameters of **A**, **B**, **D**, **A'**, **B'**, **D'** can be determined, except for the sign of $d_i$ and $d'_i$. As a matter of fact, in this case, $b$ can be deduced from (32) and then $b_i$ from (31), $d_i$ from (25), and $c_i$ from (33). The result is :

$$b_1 = \beta\sqrt{a_2a_3} \ , \qquad b_2 = \beta\sqrt{a_1a_3} \ , \qquad b_3 = \beta\sqrt{a_1a_2} \qquad (34)$$

$$d_1^2 = \beta^2\kappa\, a_1\sqrt{a_2a_3} \ , \quad d_2^2 = \beta^2\kappa\, a_2\sqrt{a_1a_3} \ , \quad d_3^2 = \beta^2\kappa\, a_3\sqrt{a_1a_2} \qquad (35)$$

$$c_1 = \gamma\sqrt{a_2a_3} \ , \quad c_2 = \gamma\sqrt{a_1a_3} \ , \quad c_3 = \gamma\sqrt{a_1a_2} \ , \quad \gamma = \beta^2(1-\kappa) - \beta \qquad (36)$$

**A'** and **B'** will be given by $\mathbf{A'} = (1-2\eta)\mathbf{A}^{-1}$, $\mathbf{B'} = (1-2\eta)\,\mathbf{B}^{-1}$ and **D'** is deduced from the elements of **B'**.

### 2.2. Case 2.2

In this case $\eta=0$. Then (30) implies $\kappa = 0$, and since **B** is positive-definite, (25) implies $d_1=d_2=d_3=0$. This condition and (4) lead to $\mathbf{D'} = 0$. Therefore, in this case, **A** and **B** are two independent matrices, $\mathbf{A'} = \mathbf{A}^{-1}$ , $\mathbf{B'} = \mathbf{B}^{-1}$ and:

$$\mathbf{D} = \mathbf{D'} = 0 \qquad (37)$$